\documentclass[%
 reprint,
 amsmath,amssymb,
 aps,
 pra,
]{revtex4-2}
\usepackage{graphicx}
\usepackage{physics}
\usepackage{xcolor}
\usepackage[]{hyperref}
\hypersetup{
    colorlinks,
    linkcolor=blue,
    citecolor=blue,
    urlcolor=blue
}

\usepackage{dcolumn}
\usepackage{bm}

\begin{document}

\preprint{APS/123-QED}

\title{A framework for separating dephasing from decoherence in\\ matter-wave Bell interferometers}
\author{S. Kannan}
\author{Y. S. Athreya}
\author{X. T. Yan}
\author{S. S. Hodgman}
\author{A. G. Truscott}
\email{andrew.truscott@anu.edu.au}
\affiliation{Department of Quantum Science and Technology, Research School of Physics, The Australian National University, Canberra, ACT 2601, Australia.
}%

\date{\today}

\begin{abstract}
Matter-wave Bell interferometers provide a sensitive probe of mass-dependent decoherence in entangled quantum systems. The degree of entanglement is obtained from the Bell-correlation amplitude of this interferometer. For observing potential mass-dependent decoherence, a reliable interpretation of any observed reduction in the Bell correlation amplitude is required, which depends on three factors: geometric dephasing, environmental decoherence, and technical dilution from source and detection statistics. In this work, we present a framework based on the Schwinger SU(2) mapping to separate these contributions into local unitaries or dissipative channels. We show that by evaluating the Bell correlation at zero interferometer path difference, it is possible to extract a source-distribution-independent Bell correlation amplitude reduction. When this framework is extended to involve atoms of different mass, we show that the known differential decoherence channels are negligible at current sensitivity. This yields a concrete bound at which a dual-species Bell interferometer would begin to signal differential decoherence beyond the known systematics, opening the way for such systems to probe new physics, such as mass-dependent decoherence mechanisms.
\end{abstract}
\maketitle
\section{Introduction}\label{sec:level1}
Atom interferometers\,\cite{RevModPhys.81.1051} harness the coherent phase accumulated along a single-atom trajectory, which can then be used for testing fundamental physics\,\cite{Schlippert2014,Tarallo2014,schuldt2015design,LMT2}. Experimental progress towards larger momentum transfer\,\cite{Kovachy2015,Asenbaum2020} and the deployment of ultracold atom interferometers in microgravity environments, which enable longer free evolution times\,\cite{PhysRevLett.110.093602,becker2018space,elliott2018nasa,Frye2021,elsen2023dual}, promise a regime for probing mass-dependent decoherence effects in quantum systems. However, single particle interferometers are insensitive to decoherence that acts symmetrically on both arms, since such processes leave the single-atom phase intact while degrading two-particle coherence.
Bell interferometers address this gap by utilizing matter waves to probe the two-particle coherence of momentum-entangled pairs through the Bell correlation amplitude\,\cite{PhysRevA.40.4277}. The Bell correlation amplitude is an observable dependent on how the joint detection probability of the two halves of the entangled Bell pairs change as the phase is changed. These correlations are affected by the loss of two-particle coherence, and hence, can be used as a complementary probe of decoherence in massive quantum systems. 

The scientific motivation is the search for mass-dependent decoherence beyond standard quantum mechanics. Several theoretical models predict decoherence rates that depend on mass and superposition size\,\cite{RevModPhys.97.015003}, such as gravitational time-dilation decoherence\,\cite{pikovski2015universal}, spontaneous collapse models\,\cite{PhysRevA.42.78,PhysRevA.40.1165,penrose1996gravity,RevModPhys.85.471}, and semiclassical gravity channels\,\cite{kafri2014classical}. Analogously to how dual-species atom interferometers isolate equivalence-principle violating differential phases\,\cite{Schlippert2014,Tarallo2014}, signals that would be invisible to single-species measurements might appear as a differential Bell-amplitude loss in dual-species tests.

The feasibility of a motional Bell interferometer with ultracold atoms has been well established theoretically\,\cite{lewisswan2015,wasak2018}, and a recent study has proposed a dual-species motional Bell test with metastable helium isotopes\,\cite{yan2025proposal}. Bell correlations have now been observed in momentum-entangled metastable helium ($^4$He$^*$) atom pairs in a matter-wave Rarity-Tapster (RT) interferometer\,\cite{thomas2022matter,athreya2026bell}. Further improvements aimed at a full CHSH Bell test with enhanced sensitivity are also underway\,\cite{PhysRevA.111.063304}. At the current experimental sensitivity in metastable helium bell interferometry\,\cite{athreya2026bell}, the predicted mass dependent decoherence signals remain many orders of magnitude beyond the experimental reach. Nevertheless, it is important to identify and bound all known differential systematic channels before any future experiment can claim a differential Bell-amplitude excess as evidence for new physics. These developments make it timely to establish a systematic framework for interpreting measured Bell-amplitude deficits, which need not arise solely from irreversible decoherence, but can also reflect reversible geometric dephasing or technical dilution from source and detection statistics.

In this work, we derive a diagnostic framework based on the Schwinger SU(2) mapping to classify the amplitude-suppressing mechanisms as local unitaries or genuinely dissipative channels. This framework establishes that evaluating the Bell correlation at zero path difference yields an exact, source-distribution-independent extraction of the Bell amplitude reduction. Furthermore, in a dual-species implementation, the ratio of Bell amplitudes at zero path difference cancels out common detection losses, making differential source mode-occupancy emerge as the dominant bounded technical systematic. Applying this framework to the $^3$He$^*$/$^4$He$^*$ system, we find that differential decoherence channels are negligible at current sensitivity, and a concrete null benchmark against which a future dual-species ratio can be tested is derived.

The paper is organized as follows: Sec.\,\ref{sec:RT_phase}  introduces the RT geometry and derives the phase-sum Bell correlation in the Schwinger representation; Sec.\,\ref{sec:ZPD} derives the zero-path-difference theorem and noise-channel classification; Sec.\,\ref{sec:dualspecies} develops the dual-species ratio, magnetic systematics, and systematic hierarchy; and Sec.\,\ref{sec:conclusion} summarizes the conclusions.

\begin{figure*}[t]
\centering
\includegraphics[width=0.95\textwidth]{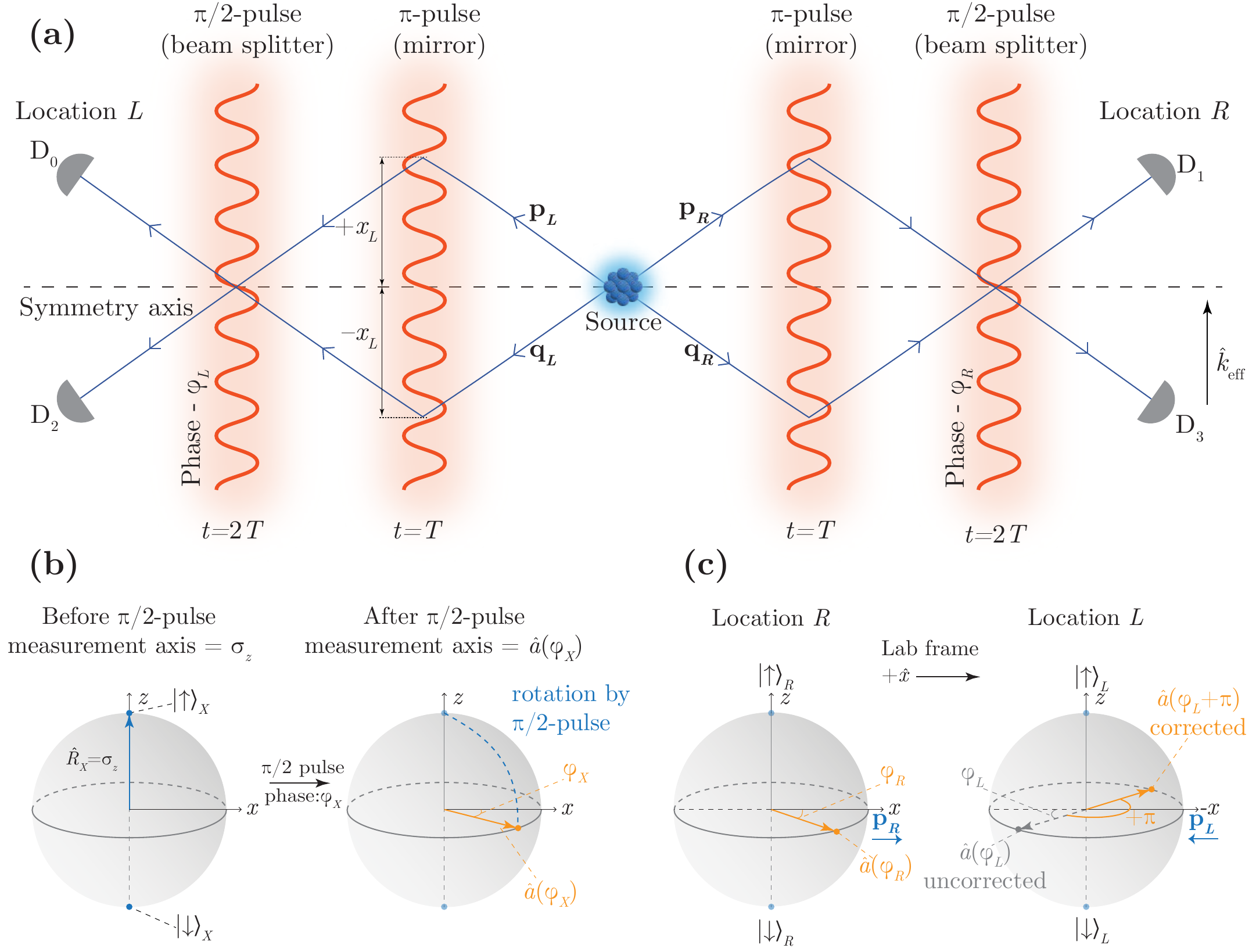}\\
\caption{(a) Schematic of the matter-wave Rarity-Tapster interferometer in the free-falling frame. Momentum-entangled pairs from the source, described by the Bell state $\vert\Psi\rangle\approx\frac{1}{\sqrt{2}}\left(\vert1_\mathbf{p},1_\mathbf{p'}\rangle+\vert1_\mathbf{q},1_\mathbf{q'}\rangle\right)$. The pairs traverse a $\pi$ mirror and $\pi/2$ beam splitter (with tunable phases $\varphi_L$ and $\varphi_R$) and arrive at locations $L$ and $R$, which comprise the momentum modes $(\mathbf{p}_L,\mathbf{q}_L)$ and $(\mathbf{p}_R,\mathbf{q}_R)$ respectively. The displacement $x_L$ is measured from the symmetry axis (dashed) to each atom's path at the $\pi$ pulse, projected onto $\hat{k}_{\rm eff}$. Detection at output ports $D_0$/$D_2$ (location $L$) and $D_1$/$D_3$ (location $R$) projects the atoms into the final momentum modes, carrying outcomes $\pm 1$.
(b) Schwinger SU(2) mapping, which maps the two-outcome momentum representation to a spin-$\tfrac{1}{2}$ system. The Bloch sphere visualizes the measurement basis, rather than the local atomic state. The poles correspond to measuring the discrete physical momentum paths $\mathbf{p}_X$ ($\ket{\uparrow}_X$) and $\mathbf{q}_X$ ($\ket{\downarrow}_X$).
In the absence of the final $\pi/2$ pulse (left), the physical detectors measure the path populations, corresponding to the local observable $\hat{R}_X=\sigma_z$. The $\pi/2$ pulse (right) coherently mixes the paths, rotating the measurement axis into the equatorial plane to a direction $\hat{a}(\varphi_X)$, where the tunable Bragg laser phase $\varphi_X$ sets the azimuthal angle of this measurement axis.
(c) Origin of the relative $\pi$ shift between measurement locations. With $|\!\uparrow\rangle_L \equiv |1_{\mathbf{p}_L}\rangle$ and $|\!\uparrow\rangle_R \equiv |1_{\mathbf{p}_R}\rangle$, the local $+x$-axis of the Bloch sphere at each location is geometrically aligned with the outward transverse direction of its respective $|\!\uparrow\rangle$ mode. Because the two atoms travel in opposite transverse directions, the local $+x$-axis at location $R$ (left sphere) naturally points to the $+\hat{x}$ in the lab frame. Conversely, the local $+x$-axis at location $L$ (right sphere) points to the left $-\hat{x}$. Expressing both measurement axes in the same lab frame therefore requires the phase correction $\varphi_L\to \varphi_L+\pi$, giving $E=-\cos\Phi$ (Eq.\,\eqref{eq:E_ideal}). Orange: corrected axis $\hat{a}(\varphi_L+\pi)$; grey dashed: initial assignment of $\hat{a}(\varphi_L)$.}
\label{fig:RT_schematic}
\end{figure*}

\section{Rarity-Tapster Bell correlations via the Schwinger SU(2) mapping}\label{sec:RT_phase}
There are several methods for generating momentum-entangled atom pairs from Bose-Einstein condensates\,\cite{PhysRevLett.99.220404,PhysRevLett.104.150402,PhysRevLett.108.260401,PhysRevA.87.063635,lewisswan2015,PhysRevLett.119.173202,PhysRevA.111.063304}. In this work, we are considering the version in Ref.\,\cite{athreya2026bell}, in which transverse momentum is conserved at the source, and detecting one atom with momentum $\mathbf{p}$ effectively projects its entangled partner into the $\mathbf{p'}$ state. In the low-mode-occupancy limit $\bar{n} \ll 1$ and only considering two mode pairs $\mathbf{p}, \mathbf{p'}$ and $\mathbf{q}, \mathbf{q'}$, the two-particle state of these scattering modes is described by a Bell state\,\cite{lewisswan2015}
\begin{equation}\label{eq1}
    \vert\Psi\rangle\approx\frac{1}{\sqrt{2}}\left(\vert1_\mathbf{p},1_\mathbf{p'}\rangle+\vert1_\mathbf{q},1_\mathbf{q'}\rangle\right),
\end{equation}
where $(\mathbf{p},\mathbf{p'})$ and $(\mathbf{q},\mathbf{q'})$ label two pairs of opposite momentum modes (see Fig.\,\ref{fig:RT_schematic}). This state is realized across two distinct halos produced by a double BEC collision\,\cite{thomas2022matter,athreya2026bell}. Each atom forms one arm of a Rarity-Tapster interferometer\,\cite{PhysRevLett.64.2495}, traversing a sequence of Bragg mirrors ($\pi$-pulse) and beam-splitter ($\pi/2$-pulse) pulses, with arm $L$ coupling the $(\mathbf{p}, \mathbf{q})$ modes and arm $R$ coupling the $(\mathbf{p'}, \mathbf{q'})$ modes (Fig.\,\ref{fig:RT_schematic}).

The RT geometry is mirror-symmetric about the source: atom\,1, with momentum $\mathbf{p}$, traverses arm $L$ at displacement $+x_L$ from the symmetry axis at the mirror-pulse, while atom 2 (momentum $\mathbf{p'}$) traverses arm $R$, also at displacement $+x_L$. Each arm traverses two identical segments, from the source to the mirror-pulse and from the mirror-pulse to the beam-splitter pulse. Consequently, the total path phase accumulated by either atom is 
\begin{equation}\label{eq:path_phase}
    \varphi^{\mathrm{path}}=\frac{2px_L}{\hbar},
\end{equation}
where $p=\mathbf{p}\cdot\hat{k}_{\rm eff}=\mathbf{p'}\cdot\hat{k}_{\rm eff}$ is the momentum projection along the effective wavevector (see Fig.\,\ref{fig:RT_schematic}), and the factor of 2 reflects the symmetry of the trajectory. Since atom\,2 traverses the corresponding upper arm with the same momentum projection and displacement, its path phase exactly matches that of atom\,1. Apart from this phase, due to free propagation, each atom obtains a phase imprinted by the beam-splitter pulse ($\varphi_L$ and $\varphi_R$). 
The total phase appears as a sum rather than a difference, as would be the case with a standard single-particle Mach–Zehnder interferometer. The physical origin of this sum structure and its impact on the Bell correlation function are established below.
At each measurement location $X\in\{L,R\}$, the final Bragg $\pi/2$ pulse coherently recombines the atomic superposition. Subsequently, detection at the output ports projects the atom into one of two final momentum modes. As illustrated in Fig.\,\ref{fig:RT_schematic}(a), for location $L$, these output modes are $\mathbf{p}_L\equiv\mathbf{p}$ and $\mathbf{q}_L\equiv\mathbf{q}$ (corresponding to detectors $D_0$ and $D_2$), while for location $R$, they are $\mathbf{p}_R\equiv\mathbf{p'}$ and $\mathbf{q}_R\equiv\mathbf{q'}$ (corresponding to detectors $D_1$ and $D_3$)
This two-outcome structure is mathematically identical to a spin-$\tfrac{1}{2}$ system. The measurement operator at the location $X$ is 
\begin{equation}
    \hat{R}_X
    =
    |1_{\mathbf{p}_X}0_{\mathbf{q}_X}\rangle
    \langle 1_{\mathbf{p}_X}0_{\mathbf{q}_X}|
    -
    |0_{\mathbf{p}_X}1_{\mathbf{q}_X}\rangle
    \langle 0_{\mathbf{p}_X}1_{\mathbf{q}_X}|,
    \label{eq:pseudospin}
\end{equation}
with eigenvalues $\pm1$ representing the two output ports ($+1$ for $D_0$/$D_1$, $-1$ for $D_2$/$D_3$; Fig.\,\ref{fig:RT_schematic}(a)).  The Schwinger mapping\,\cite{PhysRevA.33.4033} identifies these momentum-mode occupation states with the eigenstates of a pseudospin:
\begin{equation}
    |1_{\mathbf{p}_X}0_{\mathbf{q}_X}\rangle
    \longleftrightarrow |\!\uparrow\rangle_X,
    \qquad
    |0_{\mathbf{p}_X}1_{\mathbf{q}_X}\rangle
    \longleftrightarrow |\!\downarrow\rangle_X,
    \label{eq:schwinger_map}
\end{equation}
so that $\hat{R}_X$ maps to the Pauli operator $\sigma_z$. In this picture, the $\pi/2$ pulse with the tunable phase $\varphi_X$ coherently mixes the two momentum modes. Mathematically, this acts as a rotation of the measurement basis on the corresponding Bloch sphere. Specifically, $\varphi_X$ sets the azimuthal angle of the measurement axis in the equatorial plane of the Bloch sphere (Fig.\,\ref{fig:RT_schematic}(b)):
\begin{equation}
    \hat{R}_X(\varphi_X)
    \mapsto
    \hat{\sigma}\cdot\hat{a}(\varphi_X),
    \qquad
    \hat{a}(\varphi_X)
    =
    \bigl(\cos\varphi_X,\sin\varphi_X,0\bigr),
    \label{eq:bloch_rotation}
\end{equation}
where $\hat{\sigma}=(\sigma_x,\sigma_y,\sigma_z)$ is the Pauli vector. 

We emphasize that the Bloch sphere here represents the measurement operator $\hat{R}_X$ (in the Heisenberg picture), rather than the local quantum state of the individual atom\,\cite{PhysRevA.33.4033,RevModPhys.90.035005}. Because the source produces a maximally entangled bipartite state, the reduced density matrix of each atom is a maximally mixed state ($\rho = \mathbb{I}/2$). Consequently, an individual atom's local Bloch vector effectively vanishes and resides at the origin. Instead, any two-outcome projective measurement on this space is parameterized by a unit vector $\hat{a}$ indicating the measurement axis. The poles of this sphere correspond to measuring the discrete physical paths of the interferometer (momentum modes $\mathbf{p}_X$ and $\mathbf{q}_X$). The equatorial plane represents measuring a 50/50 coherent superposition of these two paths, where the azimuthal angle corresponds to the relative interferometric phase between them. Before the final $\pi/2$ pulse, detecting the atoms at the output ports corresponds to a which-path measurement along the $z$-axis. The final $\pi/2$ Bragg pulse erases this which-path information by mixing the modes, effectively rotating the measurement axis down to the equator. By tuning the Bragg laser phase $\varphi_X$, we sweep this measurement axis around the equator, directly controlling the coherent superposition projected onto the detectors. This is mathematically and conceptually identical to choosing the measurement basis vectors in a standard CHSH Bell test\,\cite{PhysRevLett.23.880}.

In the Schwinger basis, the momentum-entangled state (Eq.\,\eqref{eq1}) is represented by the Bell state
\begin{equation}
    |\Psi\rangle = \frac{1}{\sqrt{2}}
    \Bigl(|\!\uparrow\rangle_L|\!\uparrow\rangle_R
      +
      |\!\downarrow\rangle_L |\!\downarrow\rangle_R\Bigr)
    \equiv|\Phi^+\rangle.
    \label{eq:phi_plus}
\end{equation}
The correlation tensor of the state $\vert\Phi^+\rangle$ is defined as $T_{ij}=\langle\Phi^+\vert\sigma_i^{(L)}\otimes\sigma_j^{(R)}\vert\Phi^+\rangle$ (where $i,j$ denote the Pauli operators acting on the the local pseudospins at locations $L$ and $R$, respectively)\,\cite{HORODECKI1995340}, and has non-vanishing components $T_{xx} = +1$, $T_{yy} = -1$, and $T_{zz} = +1$. Physically, $T_{zz}$ represents the perfect momentum conservation of the pairs when the modes are measured directly without the final beam splitter. The transverse components $T_{xx} = +1$ and 
$T_{yy} = -1$ encode the quantum phase coherence 
of the entangled superposition. When both locations apply the $\pi/2$ beam splitter and measure along their local
$+\hat{x}$-axes, the outcomes are always identical (perfect correlation). 
When both measure along their local $+\hat{y}$-axes, the outcomes are always opposite (perfect anti-correlation). The $\pi/2$-pulses rotate the measurement axes into the equatorial plane of the Bloch sphere, giving the correlation function 
\begin{align}
    E(\varphi_L,\varphi_R)
    &=
    T_{xx}\cos\varphi_L\cos\varphi_R
    +
    T_{yy}\sin\varphi_L\sin\varphi_R\notag\\
    &=
    \cos(\varphi_L+\varphi_R).
    \label{eq:E_equatorial}
\end{align}
Because this is a bipartite Bell test with two spatially separated atoms, the measurement settings are described using two independent, local Bloch spheres, one for location $L$ and one for location $R$, as shown in Fig.\,\ref{fig:RT_schematic}(c). Equation.\,\eqref{eq:E_equatorial} assumes that the azimuthal angles on both spheres are referenced to the exact same physical direction in the lab frame. We now show that this requires a physical correction.
The natural Schwinger convention identifies $\ket{\uparrow}_L\equiv\vert 1_{\mathbf{p}}\rangle$ and $\ket{\uparrow}_R\equiv\vert 1_{\mathbf{p'}}\rangle$, consistent with the momentum conservation at the source: $\mathbf{p}$ and $\mathbf{p'}$ are always produced together as a correlated pair, so labelling them symmetrically as the respective $\ket{\uparrow}$ states is the natural choice. The azimuthal angle $\varphi_X$ is defined relative to the $x$-axis of the local Bloch sphere at each location, which we geometrically align with the physical spatial direction of the momentum mode defining $|\!\uparrow\rangle_X$. 
At location $R$, the atom travels outward to the right, so 
the state $|\!\uparrow\rangle_R = |1_{\mathbf{p}_R}\rangle$ corresponds to a momentum mode ($\mathbf{p'}$) pointing along $+\hat{x}$ in the lab frame. Thus, the local $x$-axis at $R$ points naturally along $+\hat{x}$. At location $L$, the state $|\!\uparrow\rangle_L = |1_{\mathbf{p}_L}\rangle$ 
corresponds to a momentum mode ($\mathbf{p}$) pointing along $-\hat{x}$ in the lab frame. Thus, the local $x$-axis at $L$ points along $-\hat{x}$ 
(Fig.\,\ref{fig:RT_schematic}(c)). Because these two local measurement axes point in geometrically opposite directions, 
expressing both of them relative to a single shared 
lab-frame reference requires applying a relative $\pi$ shift. Expressing both measurement axes relative to the same frame of reference (lab frame) therefore requires replacing $\varphi_L\to\varphi_L+\pi$, giving
\begin{align}
    E(\varphi_L,\varphi_R)
    &= +\cos\bigl((\varphi_L+\pi)+\varphi_R\bigr) \notag\\
    &\equiv -\cos\Phi.
    \label{eq:E_ideal}
\end{align}
This is the phase-sum structure of the correlation function, characteristic of the RT geometry. In the RT matter-wave Bell test of Ref.\,\cite{thomas2022matter}, the ideal phase-sum correlation was established using the operator formalism (treating mirror and beam-splitter pulses as rotation matrices). However, it provided no systematic way to separate geometric dephasing from irreversible decoherence in a physical experiment. Addressing this gap is the key utility of the representation developed here: imperfections that merely rotate the measurement axes are local unitaries and preserve the underlying Bell correlations, while those that contract the Bloch vector correspond to genuine decoherence (see Sec.\,\ref{sec:ZPD}). 

If experimental imperfections attenuate the ideal correlation amplitude to $E(\Phi)=-A\cos\Phi$ ($A<1$), the local-hidden-variable bound is violated when $A>1/\sqrt{2}$\,\cite{PhysRevLett.23.880}. Thus, the amplitude $A$ directly determines whether the measured correlations certify bipartite Bell nonlocality; it is not
merely a fringe-contrast parameter. In the next section, we demonstrate that evaluating the Bell correlation at zero path difference isolates the irreversible loss in Bell amplitude caused by the finite momentum spread of the source.

\section{The Zero Path Difference Scan}\label{sec:ZPD}
\subsection{Geometric dephasing and the zero path difference diagnostic}\label{sec3a}
In broadband optical interferometry, when two arms have a path difference $\delta x\neq0$, each frequency component $\omega$ accumulates a phase ($\omega \delta x/c$) before recombining, which suppresses the visibility of the fringes. This can be bypassed by operating at zero path difference (ZPD)\,\cite{griffith}, which ensures that every frequency component accumulates exactly zero phase difference between the two arms. Hence, all components are added constructively, so that the observed signal reflects the true fringe amplitude without any spectral averaging\,\cite{griffith}. 

We now extend this concept to the matter-wave regime. The Bell correlation $E(\Phi, \delta x)$ is a two-particle coincidence fringe: it oscillates sinusoidally with the interferometric phase $\Phi$, and the quantity of interest is its amplitude $A$ rather than a single-particle fringe visibility. The ZPD diagnostic is carried over directly: evaluating the coincidence fringe at $\delta x=0$ eliminates geometric phase averaging, ensuring that all momentum classes contribute in phase. 

\begin{figure}[t]
    \centering
    \includegraphics[width=\columnwidth]{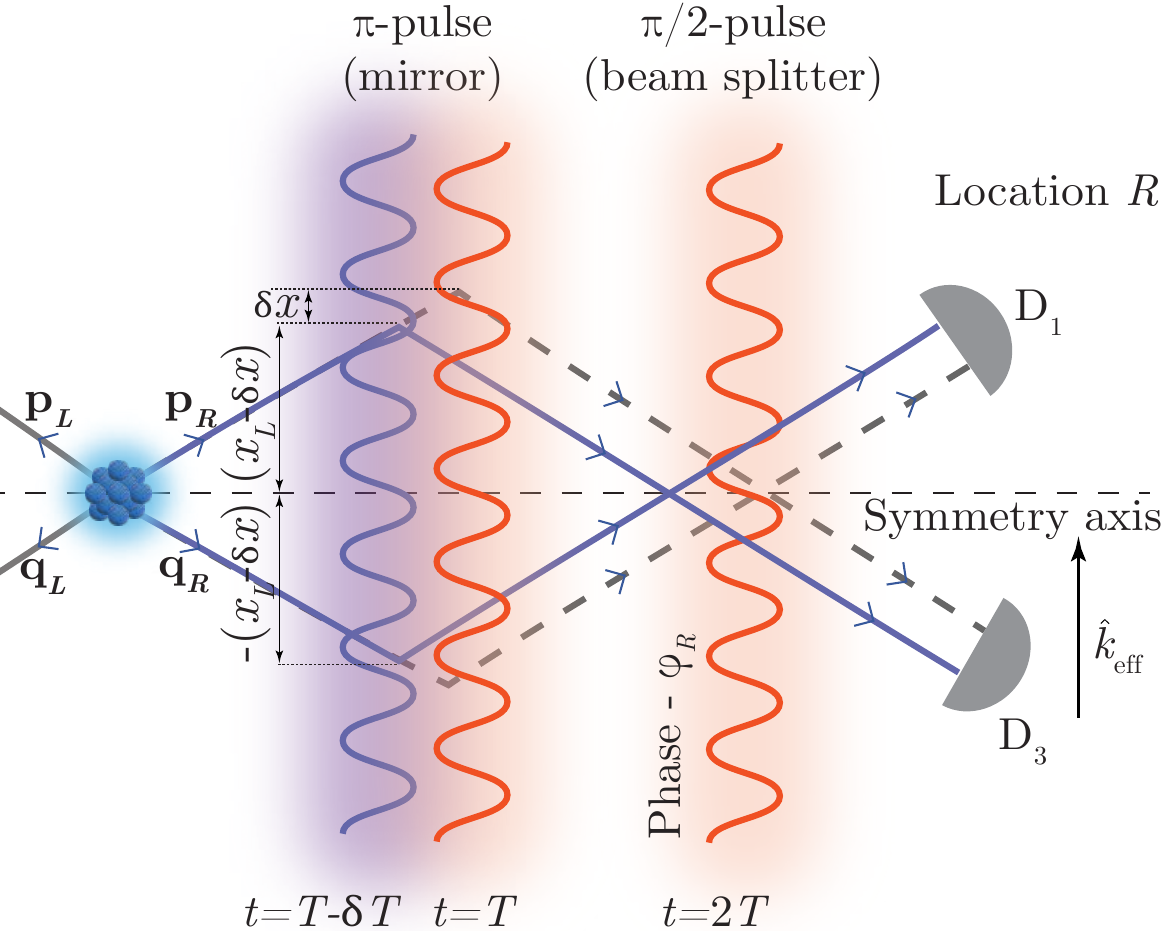}
    \caption{Right arm of the matter-wave Rarity-Tapster interferometer with a geometric path asymmetry $\delta x$. A timing shift $-\delta T$ of the mirror pulse (blue wavy line) displaces the reflection point by $\delta x$ along $\hat{k}_\text{eff}$ relative to the symmetric position at $t=T$ (red wavy line). The asymmetric path trajectories (solid blue line) traverse a displacement of $\pm(x_L-\delta x)$
 from the symmetry axis, deviating from the ideal symmetric paths (dashed blue lines).    
The asymmetry introduces an additional momentum-dependent 
geometric phase $2p_A\delta x/\hbar$ on these atoms, which varies across the source momentum distribution $f(p_A)$ and produces geometric dephasing of the Bell-correlation 
amplitude. At $\delta x = 0$ this phase vanishes simultaneously for all momentum classes, defining the zero-path-difference (ZPD) condition under which $E(\Phi,0) = -\langle A\rangle\cos\Phi$ (Eq.\,\eqref{eq:ZPD}).}
    \label{fig:ZPD}
\end{figure}

In this geometry, a path asymmetry $\delta x$ arises if the $\pi$-pulse on one arm (say, arm $R$) is applied at a position displaced by $\delta x$ along $\hat{k}_{\rm eff}$ from its symmetric value $x_L$ to $x_R=x_L-\delta x$ (Fig.\,\ref{fig:ZPD}). While the implementation in Ref.\,\cite{athreya2026bell} naturally enforces $\delta x=0$ through the use of global Bragg pulses, future experiments requiring independent control of the two arms\,\cite{yan2025proposal} may introduce a non-zero path difference. Furthermore, the source momentum distribution is never perfectly known or guaranteed to be symmetric in a real experiment. An asymmetry in the source momentum distribution will introduce a systematic error in the inferred correlation amplitude, and the ZPD condition eliminates this issue completely (Sec.\,\ref{sec3c}). Thus, the ZPD condition provides a provably source-distribution-independent extraction of the Bell amplitude reduction, which is the quantity relevant for diagnosing irreversible decoherence.

We note that while the ZPD condition removes the geometric dephasing contribution exactly, the reduced Bell-amplitude $A(p)$ can itself be momentum-dependent, because any decoherence mechanisms that depend on wave-packet separation scale with $2pt/m_A$. Away from $\delta x=0$, the momentum-dependent geometric phase and $A(p)$ are convolved within the same momentum integral and cannot be separated by visibility analysis alone. The role of the ZPD point is to remove the geometric contribution exactly (see Sec.\,\ref{sec3c}).

\subsection{General form of the correlation function}\label{sec3b}
We now examine the general form of the correlation function in the presence of geometric asymmetry. The total path phase accumulated by an atom $B$ on arm $R$ (assuming its partner is atom $A$ on arm $L$) is
\begin{align}
    \varphi^{\rm path}_B
    &= \frac{2\,p_B\,x_R}{\hbar}
    = \frac{2\,p_A(x_L-\delta x)}{\hbar}\notag\\
    &= \frac{2\,p_A x_L}{\hbar} - \frac{2\,p_A\delta x}{\hbar},
    \label{eq:phase_asym}
\end{align}
where $p_B$ and $p_A$ are the vertical momentum component of the corresponding atoms in the two arms.  The first component of the path phase matches that of the atom on arm $L$, and the second term $2\,p_A\delta x/\hbar$ represents the additional geometric phase from arm asymmetry. Because it is proportional to $p_A$, it varies across the source momentum distribution, causing the geometric dephasing described in Sec.\,\ref{sec3a}.

In the Schwinger picture, this additional phase acts as a local equatorial rotation on the Bloch sphere at location $R$ by angle $2\,p_A\delta x/\hbar$ (see Fig.\,\ref{fig:RT_schematic}(b)). Consequently, the Bell correlation for a fixed momentum class $p_A$ becomes:
\begin{equation}
    E(\Phi,\delta x\,|\,p_A) 
    = -A(p_A)\cos\!\left(\Phi - \frac{2p_A\delta x}{\hbar}\right),
    \label{eq:E_fixed_p}
\end{equation}
where $A(p_A)$ accounts for the momentum-dependent Bell-amplitude reduction. Averaging over the source momentum distribution $f(p_A)$ gives
\begin{equation}
    E(\Phi,\delta x) = -\int_{-\infty}^{\infty} f(p_A)\,A(p_A)
    \cos\!\left(\Phi -\frac{2p_A\delta x}{\hbar}\right)dp_A.
    \label{eq:E_general}
\end{equation}
Note that each momentum class $p_A$ is perfectly correlated with its partner atom by momentum conservation at the source; the dephasing arises because the geometric phase $2p_A\delta x/\hbar$ takes a different value for each correctly-matched pair, causing their fringes to add with relative phase offsets when integrated over $f(p_A)$.
This result highlights that the interpretation of a measured Bell-amplitude deficit is nontrivial away from the ZPD. Even if $f(p_A)$ were independently known, extracting the underlying decoherence amplitude away from the ZPD point would require either an error-prone numerical deconvolution of the data or an assumed theoretical model for the functional form of $A(p_A)$.

\subsection{Extraction of the momentum-averaged Bell-amplitude reduction at ZPD}\label{sec3c}
The significance of the ZPD condition is evident from Eq.\,\eqref{eq:E_general}. At $\delta x=0$, the cosine argument reduces to $\Phi$ for all $p_A$, so the phase factor is removed from the momentum average. In this limit, the Bell correlation becomes
\begin{equation}
    E(\Phi,0) = -\cos\Phi\int_{-\infty}^{\infty}f(p_A)\,A(p_A)\,dp_A 
    \equiv -\langle A\rangle \cos\Phi. 
    \label{eq:ZPD}
\end{equation}
Note that $\langle A\rangle$ is the quantity that directly determines whether the measured correlations certify Bell nonlocality, and at the ZPD this value is extracted without any model of $f(p_A)$. 

\begin{figure}[t]
    \centering
    \includegraphics[width=\columnwidth]{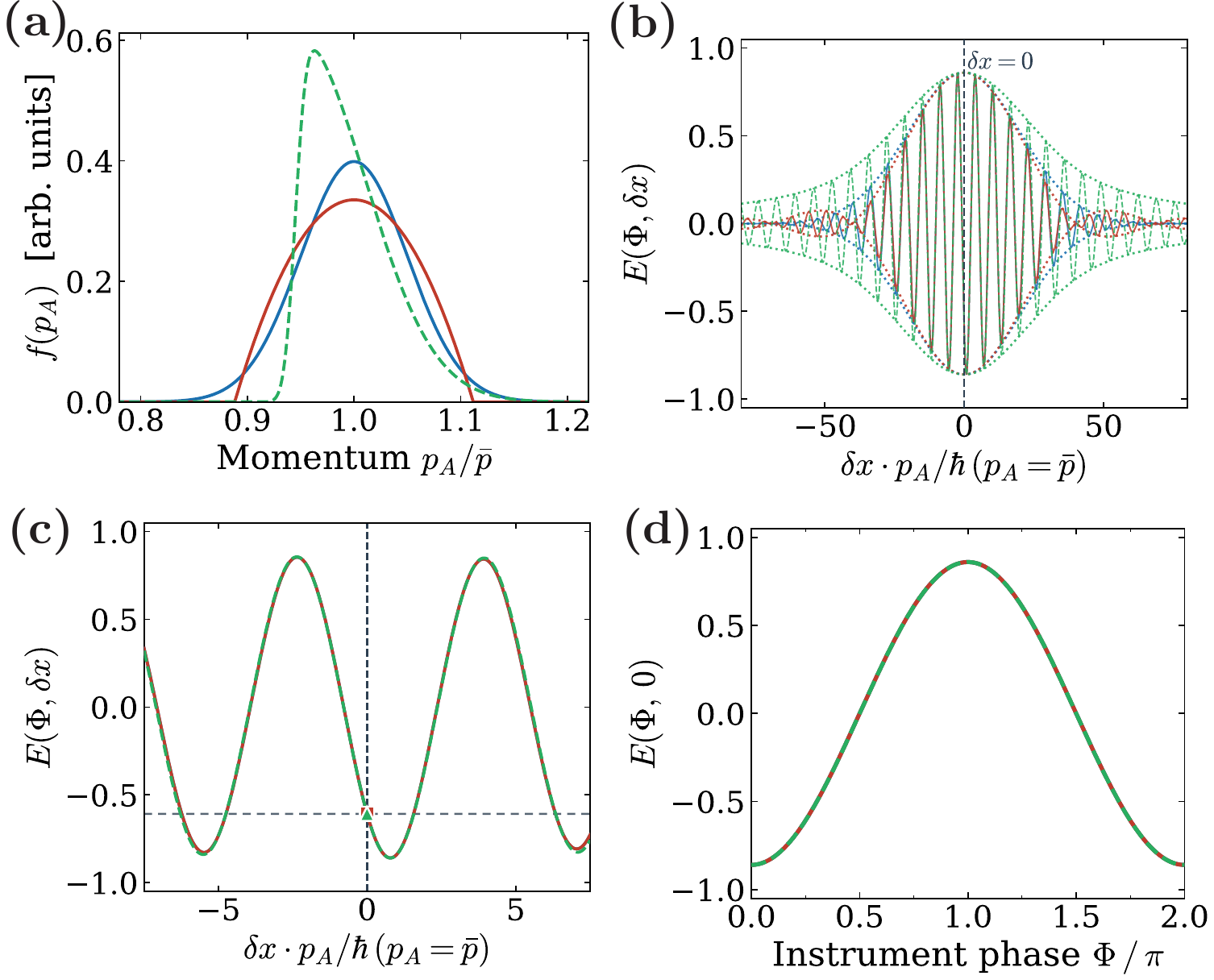}
    \caption{Separation of geometric dephasing and Bell-amplitude reduction in a $\delta x$ scan. (a) Representative source momentum distributions $f(p_A)$: Gaussian
    (blue), Thomas-Fermi (red), and asymmetric (green dashed).
    (b) Corresponding Bell-correlation interferograms $E(\Phi,\delta x)$ bounded by the geometric envelopes $\langle A\rangle\,|\widetilde{F}(\delta x/\hbar)|$, while (c) shows the region near $\delta x=0$. Despite different fringe phases and
    envelope shapes away from zero, all three distributions yield the
    same Bell-correlation value at the ZPD point (Eq.\,\eqref{eq:ZPD}), confirming the exact, source-distribution-independent extraction of $\langle A\rangle$.
    (d) $\Phi$ scan at fixed $\delta x=0$: all three cases collapse onto
    the same sinusoid $-\langle A\rangle\cos\Phi$, from which the
    Bell-amplitude reduction $\langle A\rangle$ is extracted directly. 
    Simulation parameters: $\Delta p/\bar{p}=0.05$, $\langle A\rangle
    =0.86$, and $\Phi=\pi/4$.}
    \label{fig:dx_scan}
\end{figure}

Experimentally, identifying the ZPD point requires maximizing the measured Bell correlation amplitude $\langle A\rangle$ as $\delta x$ varies. To illustrate this, we consider the regime where the fractional momentum spread is small, $\Delta p/\bar{p} \ll 1$ (with $\Delta p$ being the momentum width and $\bar{p}$ the mean momentum) and $A(p_A)$ is approximately constant across $f(p_A)$. The correlation function then takes the form:
\begin{equation}
    E(\Phi, \delta x) \approx -\langle A\rangle\,|\tilde{F}(\delta x/\hbar)|
    \cos\!\left(\Phi + \arg\tilde{F}(\delta x/\hbar)\right),
    \label{eq:E_approx}
\end{equation}
where
\begin{equation}
\tilde{F}(\delta x/\hbar) = \int_{-\infty}^{\infty} f(p_A)\,e^{-i2p_A\delta x/\hbar}\,dp_A
    \label{eq:triangle}
\end{equation}
is the geometric dephasing envelope. Since $f(p_A)$ is normalized, the triangle inequality ($|\tilde{F}(k)|\leq\tilde{F}(0) = 1$ for all $k$), ensures that the envelope maximum occurs at $\delta x = 0$ for any source distribution (Fig.\,\ref{fig:dx_scan}).

The ZPD point is located by sweeping the geometric asymmetry $\delta x$ across the characteristic coherence length of the dephasing envelope ($|\delta x|\lesssim \hbar/\Delta p$).
In practice, $\delta x$ can be tuned via the relative timing of the mirror pulse on arm $R$, while the interferometric phase $\Phi$ is scanned at each step to record $E(\Phi,\delta x)$. A sinusoidal fit to the measured $E(\Phi,\delta x)$ extracts the local correlation amplitude $A(\delta x)=\langle A\rangle|\tilde{F}(\delta x/\hbar)|$, which reaches its maximum at the ZPD.

Beyond the theoretical motivation, the ZPD point is highly robust against realistic pulse-timing imperfections. In practice, the $\pi$-pulse timing on one arm sets the asymmetry $\delta x$, meaning any timing uncertainty $\delta t$ directly introduces a spatial uncertainty $\delta x \sim v_r \delta t$, where $v_r$ is the mean recoil velocity. Away from the ZPD, this produces a systematic error in the extracted amplitude because the measured correlation sits on a sloping region of the dephasing envelope, $|\tilde{F}(\delta x/\hbar)|$. 
Additionally, because atomic momentum is reconstructed from detector arrival times\,\cite{thomas2022matter,athreya2026bell}, a timing error $\delta t$ introduces a momentum error $\delta p_B \sim m v_r \delta t / T$ for the partner atom. This corrupts the $p_A = p_B$ coincidence condition, admitting accidental counts while excluding true pairs. 
Operating at the ZPD provides a robust safeguard against the first issue. Since the dephasing envelope is stationary at its maximum ($d|\tilde{F}|/d(\delta x)|_{\delta x=0} = 0$), small timing uncertainties around the ZPD point perturb $\langle A \rangle$ only to second order. While resolving the momentum reconstruction issue does require a precise, independent calibration of $\delta t$, this is exactly the calibration already achieved when locating the ZPD point via the $A(\delta x)$ scan. In this way, the two procedures are naturally unified.

\subsection{Reversible dephasing and irreversible loss of Bell-amplitude}
The distinction between reversible dephasing and irreversible decoherence is commonly used in other interferometric settings, such as inhomogeneous broadening in optics\,\cite{allen2012optical} and $T_2^*$ versus $T_2$ in magnetic resonance\,\cite{PhysRev.80.580}. In our bipartite setting, the distinction manifests in a clear algebraic form.

When $\delta x\neq0$, the fringe from momentum class $p_A$ is shifted in phase relative to the fringe from $p_A'$ by $2(p_A - p_A')\delta x/\hbar$ (Eq.\,\ref{eq:phase_asym}). When averaged over the source momentum distribution $f(p_A)$, these offsets result in a reduced Bell correlation amplitude. This geometric dephasing is precisely analogous to inhomogeneous broadening in optics, where an ensemble average over varying frequencies washes out the fringe contrast\,\cite{RevModPhys.75.715}. The role of the ZPD point is analogous to that of a spin echo in magnetic resonance\,\cite{PhysRev.80.580}. Just as a $\pi$ pulse refocuses the  inhomogeneous dephasing caused by a distribution of local precession frequencies, stripping away the $T_2^*$ envelope to reveal the true $T_2$ decay\,\cite{PhysRev.80.580}, operating at $\delta x=0$ refocuses the momentum-dependent geometric phases to reveal the true Bell-amplitude reduction, $\langle A\rangle$.

In contrast, irreversible processes reduce the Bloch vector length and cannot be compensated for by setting $\delta x=0$. Irreversible processes produce phases or amplitude losses that vary stochastically from shot to shot, so that no single parameter adjustment can recover the lost coherence. The correspondence with magnetic resonance is
\begin{align}
    \Delta B&\longleftrightarrow\Delta p,\notag\\
    \text{local precession offset}&\longleftrightarrow\frac{2p_A\delta x}{\hbar},\notag\\
    T_2^* \text{ envelope}&\longleftrightarrow\left|\widetilde{F}\!\left(\frac{\delta x}{\hbar}\right)\right|.
    \label{eq:nmr_mapping}
\end{align}
Here $\Delta B$ is the spread of local magnetic field offsets across the spin ensemble and $T_2^*$ is the observed transverse relaxation time containing both reversible inhomogeneous dephasing and irreversible loss. 
In this analogy, $\langle A\rangle$ measured away from $\delta x = 0$ 
corresponds to the $T_2^*$ amplitude and $\langle A\rangle$ extracted at $\delta x = 0$ 
via Eq.\,\eqref{eq:ZPD} corresponds to the $T_2$ (intrinsic decoherence time that survives the spin echo refocusing) amplitude, isolating the irreversible loss alone.
\subsection{Classification of experimental imperfections}\label{sec3e}
Within the Schwinger SU(2) framework, experimental imperfections are classified as local unitaries (preserving Bell amplitude) or as non-unitary processes (degrading Bell amplitude). At the level of the measured signal, we further distinguish between local unitary phases that are fixed and shot-invariant, and therefore absorbed into the fitted phase offset without reducing $\langle A\rangle$, and stochastic shot-to-shot fluctuations that survive ensemble averaging and produce a genuine reduction of the Bell-correlation amplitude. More details are given in Appendix.\,\ref{app:A}.

At the ZPD point, we effectively remove momentum-dependent equatorial phases due to the path asymmetry. Any residual fixed offsets are handled by the phase fit $\Phi_0$. We identify four distinct channels that are relevant to the present system.

\noindent\textit{Shot-to-shot phase jitter} (e.g. Bragg phase or magnetic field noise) produces stochastic dephasing described by a local jump operator $L_X =\sqrt{\Gamma_\phi}\,\sigma_z^{(X)}$ (see Appendix.\,\ref{app:A}), where $\Gamma_\phi$ is the dephasing rate and $\sigma_z^{(X)}$ is the Pauli operator acting on the pseudospin at location $X$.
This process preserves $T_{zz} = +1$ but decays the transverse correlations as $T_{xx} = -T_{yy} = e^{-2\Gamma_\phi t}$ (where $t$ is the elapsed measurement time), reducing $\langle A\rangle$ upon ensemble averaging. Shot-to-shot magnetic noise falls into this category, whereas static magnetic gradients, being coherent and shot-invariant, are local unitaries and hence do not.
\\
\noindent\textit{Atom loss:} Processes such as background-gas scattering and resonant light scattering ejects atoms from the relevant momentum modes into undetected states, destroying all correlations. In the low-mode occupancy regime $\bar{n} \ll 1$, the dominant effect is uniform atom loss from the coincidence dataset, which does not reduce $\langle A\rangle$ in normalized counts, although at finite $\bar{n}$ there will also be a small degradation of $\langle A\rangle$. 
\\
\noindent\textit{Magnetic field curvature} produces a momentum-dependent residual phase 
$\delta\phi_{\rm curv}(p_A) \propto p_A$ that survives both the RT mirror-symmetry cancellation and the ZPD condition (Appendix.\,\ref{app:d}). Although this phase is coherent rather than stochastic, it is not removable by any single protocol adjustment and therefore constitutes an effective irreducible reduction of $\langle A\rangle$.
\\
\noindent\textit{Source statistics}
Higher-order Fock-state contributions from the two-mode squeezed-vacuum source reduce the measured Bell amplitude to $\langle A\rangle_{\rm stat} = (1+\bar{n})/(1+3\bar{n})$ 
\cite{lewisswan2015,wasak2018}, giving $\langle A\rangle_{\rm stat} \approx 0.937$ for 
$\bar{n} \approx 0.035$ used in Ref.\,\cite{athreya2026bell}. This is not technically a decoherence channel, but a consequence of the multi-mode source statistics. It could be reduced by operating at lower mode occupancy, although data acquisition rates make this challenging\,\cite{athreya2026bell}.

\section{Differential Decoherence and Systematic Bounds in Dual-Species Bell tests}\label{sec:dualspecies}
\subsection{The dual-species ratio}\label{sec4a}
While the single-species setup in Ref.\,\cite{athreya2026bell} utilizes entangled $^4$He$^*$-$^4$He$^*$ pairs, characterized by the correlation amplitude $A_4$, the dual-species experiment proposed in Ref.\,\cite{yan2025proposal} would produce an inter-species entangled state with $^3$He$^*$-$^4$He$^*$ pairs
\begin{equation}
    |\Psi\rangle \approx \frac{1}{\sqrt{2}}
    \left(|1_\mathbf{p}^{(3)}, 1_\mathbf{p'}^{(4)}\rangle 
    + |1_\mathbf{q}^{(3)}, 1_\mathbf{q'}^{(4)}\rangle\right),
    \label{eq:dual_bell_state}
\end{equation}
where the superscript (3/4) represents the atomic species ($^3$He$^*$/$^4$He$^*$).
In this bipartite system, any decoherence affecting either species will necessarily reduce the joint correlation $E_{3-4}(\Phi)$, which is constructed from $^3$He$^*$-$^4$He$^*$ coincidence measurements. By comparing the dual-species amplitude $A_{3-4}$ with the single-species amplitude $A_4$ at the ZPD, their ratio provides a robust diagnostic for detecting species-dependent decoherence.

At the ZPD point, any common multiplicative detection losses cancel out, ensuring they do not bias the dual-species comparison. For metastable helium measured on a microchannel plate\,\cite{Manning:10}, the single-atom detection efficiencies for $^3$He$^*$ and $^4$He$^*$ are identical for our purposes\,\cite{PhysRevLett.134.223001}, and can be anywhere between $7-50\%$ depending on the setup\,\cite{PhysRevLett.105.190402,hoendervanger2013influence,PhysRevA.97.063601,PhysRevA.110.063324,PRXQuantum.5.040324,h7ws-g9z2}. 

Let $A_\textrm{local}$ be the Bell correlation amplitude reduction common to both scenarios and let $\delta A>0$ be the additional reduction suffered by the dual-species correlation amplitude due to species dependent decoherence channels. At ZPD, the ratio of their amplitudes is
\begin{equation}
    \frac{A_{3-4}}{A_4} = \frac{\langle A_{\rm local} - \delta A\rangle}{\langle A_{\rm local}\rangle} = 1-\frac{\langle\delta A\rangle}{\langle A_{\rm local}\rangle}.
    \label{eq:ratio_main}
\end{equation}
A separate reduction of the absolute Bell amplitude arises from the thermal statistics of the two-mode squeezed-vacuum source\,\cite{lewisswan2015,wasak2018}. The source statistics factor cancels the ratio $A_{3-4}/A_4$ only if the mean occupancies are the same between the two configurations. Including the source-statistics correction, the measured Bell amplitudes are
\begin{align}
    A_4 &= \langle A_{\rm local}\rangle\,\frac{1+\bar{n}_4}{1+3\bar{n}_4},
\label{eq:A4_full}\\
    A_{3-4} &= \langle A_{\rm local} - \delta A\rangle\,\frac{1+\bar{n}_{3-4}}{1+3\bar{n}_{3-4}},\label{eq:A34_full}
\end{align}
where $\bar{n}_4$ and $\bar{n}_{3-4}$ are the mean mode occupancies in the single-species and dual-species configurations respectively. We note that potential corrections arising from the fermionic statistics of $^3$He$^*$ are omitted here and remain beyond the scope of this work. Although this is an oversimplification, the key point is that differential source occupancy can introduce a systematic shift in the resulting ratio.

Expanding the ratio $A_{3-4}/A_4$ to first order in $\delta A$ and $\Delta\bar{n} = \bar{n}_{3-4} - \bar{n}_4$ gives
\begin{equation}
    \frac{A_{3-4}}{A_4} \approx 1
    - \frac{\langle\delta A\rangle}
           {\langle A_{\rm local}\rangle}
    - \frac{2\,\Delta\bar{n}}
           {(1+\bar{n})(1+3\bar{n})},
    \label{eq:ratio_expanded_main}
\end{equation}
where $\bar{n}$ is the average mode occupancy ($\bar{n}=\tfrac{1}{2}(\bar{n}_{3-4} + \bar{n}_4)$), and corrections of order $\mathcal{O}(\delta A^2,\, \delta A\,\Delta\bar{n},\, \Delta\bar{n}^2)$ neglected.
The corresponding contribution from source occupancy imbalance is
\begin{equation}
    \langle\delta A\rangle_{\Delta\bar{n}}
    = \frac{2\,\Delta\bar{n}}
           {(1+\bar{n})(1+3\bar{n})}
    \,\langle A_{\rm local}\rangle.
    \label{eq:accidental_main}
\end{equation}
Equation~\eqref{eq:accidental_main} identifies differential source occupancy as the leading bounded technical systematic that survives the dual-species ratio. Details of the source-statistics derivation are given in Appendix~\ref{app:c}.

\subsection{Environmental and optical dissipation}\label{sec4c}
Unlike geometric phase effects, dissipative channels produce genuine nonunitary Bell-amplitude loss and are not removed by the ZPD protocol or by RT mirror symmetry. For the dual-species helium platform\,\cite{yan2025proposal}, the two leading candidates are background-gas collisions and spontaneous emission during the Bragg sequence. Both have shown to be negligible for the $^4$He$^*$ setup\,\cite{athreya2026bell}, and would be expected to also be negligible for the dual species setup at current and foreseeable sensitivity.
\subsection{Experimentally observed amplitude reduction}\label{sec4d}
\begin{figure}[t]
    \centering
    \includegraphics[width=\columnwidth]{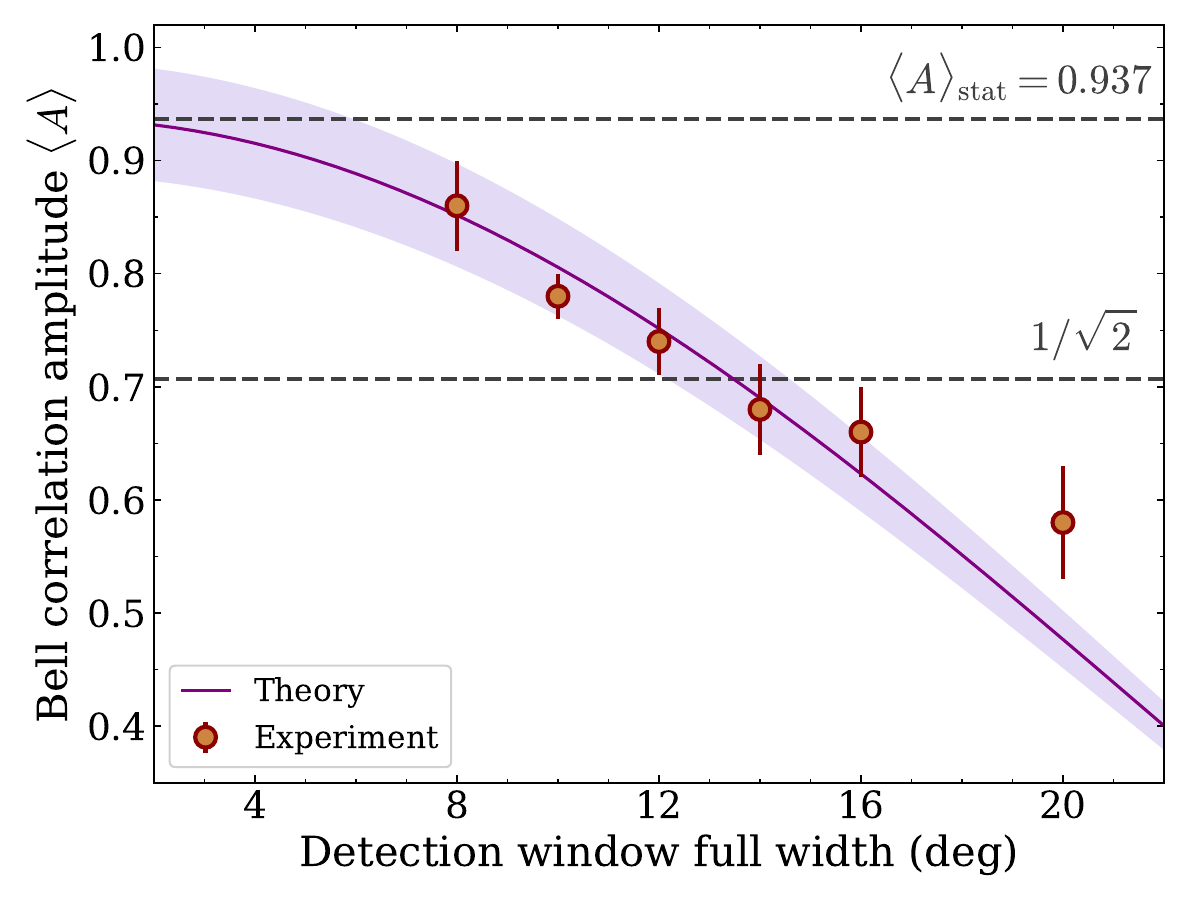}
    \caption{Bell correlation amplitude $\langle A\rangle$ as a function of the full width of the angular detection window ($W$). Filled circles are extracted from a sinusoidal fit to the experimental data in \,\cite{athreya2026bell}. The solid curve shows the theoretical prediction from Eq.\,\eqref{eq:sinc_reduction}, with $\Delta\phi_\text{window}=1.5\times W/8$, calibrated from the $\pm4^\circ$ window. The purple shaded region indicates the uncertainty arising from mode-occupancy. The theory reproduces the data well for windows up to 14$^\circ$, beyond which the linear approximation breaks down.}
\label{fig:exp-comp}
\end{figure}
All known channels of decoherence are far below current experimental sensitivity in the metastable helium platform. However, the experimentally observed single-species $^4$He$^*$ Bell amplitude\,\cite{athreya2026bell}, $\langle A\rangle = 0.86\pm0.03$, lies below the 
source-statistics-limited expectation $\langle A\rangle_{\rm stat} = (1+\bar{n})/(1+3\bar{n}) \approx 0.937$ by $0.077$, corresponding to a $\sim2.6\sigma$ deficit. We now present a physical explanation for this discrepancy.

\noindent\emph{Detection window phase variation:}
The $s$-wave scattering halo is spherical, and atoms detected outside the equatorial plane have momentum vectors pointing in slightly different directions\,\cite{athreya2026bell}. An atom at polar angle $\theta$ from the equator has a momentum component $p_\parallel(\theta) = 
\bar{p}\sin\theta \approx \bar{p}\theta$ along the effective wavevector $\hat{k}_{\rm eff}$, which introduces the path phase (Eq.\,\eqref{eq:path_phase})
\begin{equation}
    \varphi^{\rm path}(\theta) 
    = \frac{2\bar{p}\,x_L\,\theta}{\hbar}.
    \label{eq:angular_phase}
\end{equation}
Additionally, each atom experiences a Doppler detuning $\delta(\theta) = k_{\rm eff}\bar{p}\theta/m$ that modifies the effective Bragg phase imprinted at each mirror and beamsplitter. Together these produce a total interferometric phase variation of approximately 
$ 1.5$\,rad ($\Delta\phi_{\rm window}\approx1.5\,$rad) across the $\pm4^\circ$ detection window averaged over in\,\cite{athreya2026bell}.

This angular phase variation is distinct from (and additional to) the radial momentum spread of the source addressed by the ZPD theorem. It arises from the angular distribution of detected momenta on the halo and hence will be present even at $\delta x = 0$. 

The Bell correlation, averaged over atoms distributed uniformly across the detection window, becomes
\begin{align}
    E(\Phi) &= -\langle A\rangle_{\rm stat}
    \cdot\frac{1}{2\Theta}
    \int_{-\Theta}^{+\Theta}
    \cos\!\left(\Phi + 
    \frac{\Delta\phi_{\rm window}}{2}
    \cdot\frac{\theta}{\Theta}\right)d\theta
    \notag\\
    &= -\langle A\rangle_{\rm stat}\cos\Phi
    \cdot\frac{\sin(\Delta\phi_{\rm window}/2)}
               {\Delta\phi_{\rm window}/2},
    \label{eq:window_avg}
\end{align}
where $\Theta$ is the half-width of the detection window. The measured Bell amplitude is therefore reduced by the sinc factor
\begin{equation}
    \langle A\rangle_{\rm meas}
    = \langle A\rangle_{\rm stat}\cdot
    \left|
    \frac{\sin(\Delta\phi_{\rm window}/2)}
         {\Delta\phi_{\rm window}/2}
    \right|.
    \label{eq:sinc_reduction}
\end{equation}
For $\Delta\phi_{\rm window} = 1.5$\,rad, this gives a reduction factor of $\sin(0.75)/0.75 \approx 0.908$. Thus the expected correlation amplitude is:
\begin{equation}
    \langle A\rangle_{\rm exp}
    \approx 0.937\times0.908\approx0.850,
    \label{eq:A_pred_combined}
\end{equation}
in good agreement with the observed value $0.86\pm0.03$. This deficit is in principle reducible by narrowing the detection window at the cost of signal-to-noise ratio, as discussed in Ref.\,\cite{athreya2026bell}. In Fig.\,\ref{fig:exp-comp} we plot the data from\,\cite{athreya2026bell} alongside our model (Eqs.\,\eqref{eq:window_avg}-\eqref{eq:sinc_reduction}) for different detection window widths.  The model agrees well with the data, showing that the reduction in $\langle A\rangle$ observed in \cite{athreya2026bell} can be fully explained by mode occupancy statistics and finite detection window, rather than known or anomalous decoherence mechanisms. Furthermore, the finite-width of atomic wavefunctions\,\cite{PhysRevA.88.013617} contributes an additional reduction of $\lesssim 0.01$, well below the current statistical precision. This boundary effect arises because the atoms are not localized to a point (the wavepackets have finite width), and only a fraction of the wavepackets near the boundary of the detection window is captured, which reduces their detection probability.

\emph{Null benchmark for the dual-species ratio:}
Our null hypothesis is based on standard quantum mechanics and the decoherence channels that we have discussed in the preceding sections. Within this framework, any observed discrepancy in the ratio $A_{3-4}/A_4$ would point to new mass-dependent decoherence mechanisms that are not accounted for by standard atomic physics. 

\begin{figure}[t]
    \centering
    \includegraphics[width=\columnwidth]{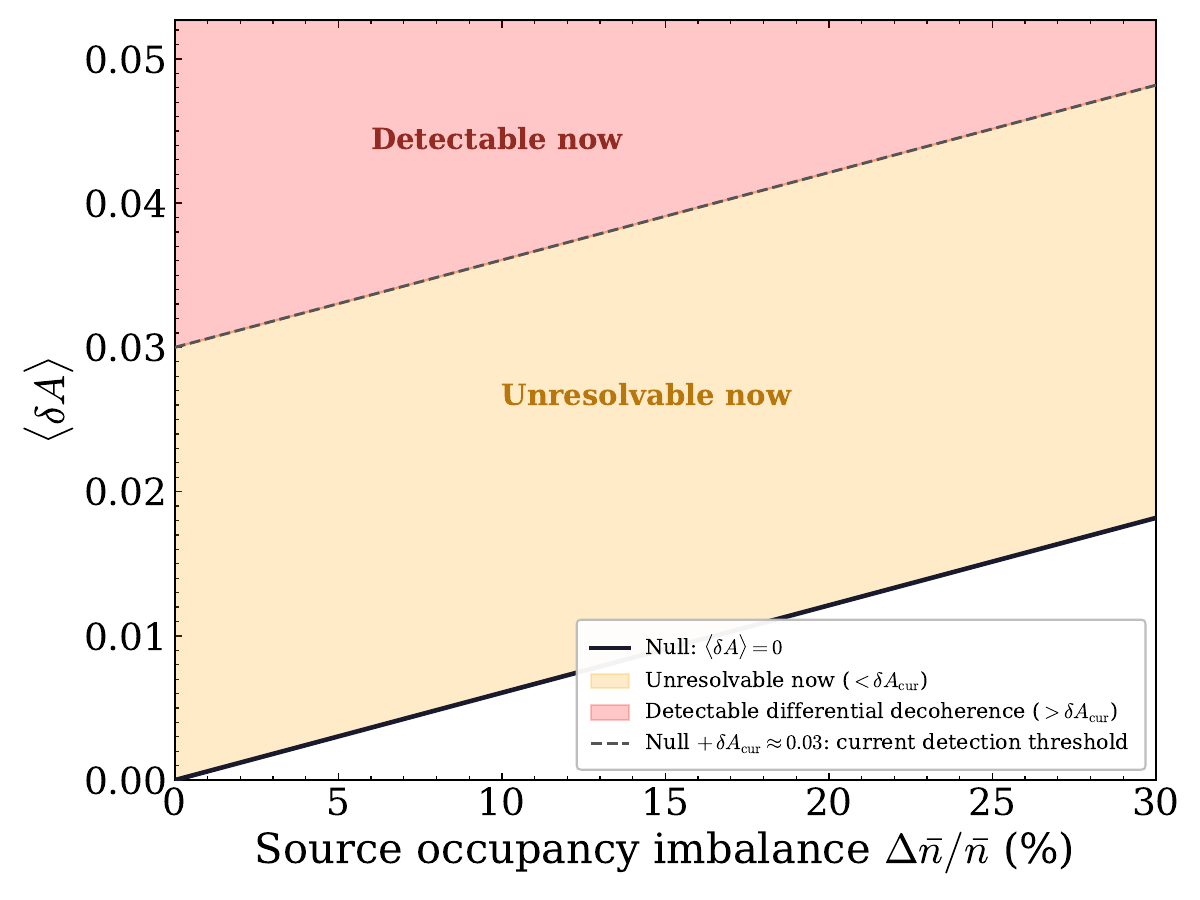}
    \caption{Null-hypothesis benchmark for the dual-species Bell-amplitude
comparison. The $y$-axis is the measured quantity $\langle\delta A\rangle$ (Eq.\,\eqref{eq:ratio_expanded_main}). The solid black curve gives the null prediction at which $\langle\delta A\rangle=0$, accounting for finite source occupancy. A measured point cannot lie below the null curve; any vertical excess above it is the genuine differential decoherence signal $\langle\delta A\rangle$. The orange band marks the region where $\langle\delta A\rangle$ is present but below the current statistical uncertainty $\delta A_{\rm cur}\approx 0.03$ (grey dashed line) and therefore unresolvable now; the red band marks the region where $\langle\delta A\rangle$ is detectable with the current experiment.}
\label{fig:null_benchmark}
\end{figure}

Under this null hypothesis, the deviation from unity is governed by two bounded technical systematics. First, because the Doppler detuning phase variation across the halo ($\delta(\theta)\propto1/m$) depends on atomic mass, the finite-window amplitude reduction (Eq.\,\eqref{eq:sinc_reduction}) will differ slightly between the single and dual-species configurations. This introduces a calculable, purely kinematic differential shift. Once this known detection-window effect is factored out, the surviving deviation arises solely from the imbalance in source occupancy (Eq.\,\eqref{eq:ratio_expanded_main}):
\begin{equation}
    \frac{A_{3-4}}{A_4}\Big|_{\rm null}
    \sim 1 - 
    \frac{2\,\Delta\bar{n}}
         {(1+\bar{n})(1+3\bar{n})}.
    \label{eq:null_benchmark}
\end{equation}
As shown in Fig.\,\ref{fig:null_benchmark} the solid black line establishes a baseline where the amplitude deficit grows linearly with source occupancy imbalance. Because real differential decoherence only drives the correlation amplitude down further, valid measurements must appear on or above this line. Therefore, if a measurement lands in the red shaded region, exceeding both the baseline and the margin of error, it proves the existence of genuine decoherence signal.

\subsection{Generality of the framework and scaling to LMT and microgravity platforms}\label{sec4e}
Although the quantitative estimates in this work are tailored to the $^3$He$^*$/$^4$He$^*$ platform, the underlying diagnostic framework is more general. The ZPD extraction, the SU(2) classification of imperfections, and the symmetry-based cancellation of linear magnetic gradients apply to mirror-symmetric momentum-entangled Bell interferometers in which the measured phase enters as a sum ($\Phi=\varphi_L+\varphi_R$). For other systems (such as momentum-entangled pairs generated from Rubidium or Potassium BECs\,\cite{doi:10.1126/science.aao1850,RevModPhys.90.035005,PhysRevLett.127.140402}), the systematic analysis remains the same; we only need to update a few physical parameters.

The null benchmark established in Sec.\,\ref{sec4d} motivates a natural question: what experimental improvements are needed to reach sensitivity to the gravitational and 
collapse-model decoherence signals that the dual-species ratio is designed to probe? Both 
Pikovski-type gravitational decoherence\,\cite{pikovski2015universal} and spontaneous 
collapse models\,\cite{PhysRevA.42.78,PhysRevA.40.1165,penrose1996gravity} predict decoherence rates that scale with the spatial superposition size $\Delta x$. In 
the RT geometry, $\Delta x \sim \hbar k_{\rm eff} T/m$ is set by the recoil momentum and the free-evolution time. 

Two complementary routes exist to increase $\Delta x$ and thereby the sensitivity to these signals: large-momentum-transfer (LMT) beam splitters, which increase $k_{\rm eff}$ by using multiple Bragg photon kicks\,\cite{PhysRevLett.100.180405, PhysRevLett.107.130403,Kovachy2015,Asenbaum2020}; and the move to microgravity or space-based platforms, which allow much longer free-evolution times $T$\,\cite{PhysRevLett.110.093602,becker2018space,elliott2018nasa,Frye2021,elsen2023dual}. Both of these change the experimental requirements for the diagnostic framework and the ZPD theorem and systematic classification developed here becomes vital.  

First, in large-momentum-transfer (LMT) implementations, the geometric dephasing envelope $|\tilde{F}(\delta x/\hbar)|$ narrows as the momentum splitting increases. Even small uncontrolled arm asymmetries (e.g. imperfect pulse timing, vibrations, etc.) produce significant geometric dephasing and 
Bell-amplitude suppression. Precisely locating and maintaining the ZPD point is therefore an essential experimental requirement.

Second, in microgravity or space-based platforms, the longer free-evolution time $T$ has two consequences. The arm separation at the mirror pulse grows as $v_r T$, eventually becoming large enough that independent beam-addressability of the two arms is both highly practical and operationally necessary. Once independent arm addressability is implemented, controlled or uncontrolled path asymmetries $\delta x \neq 0$ can arise, and the ZPD framework is needed to diagnose and correct for them.


\section{Conclusion}\label{sec:conclusion}
We have developed a diagnostic framework for interpreting Bell-correlation amplitudes in matter-wave interferometers in the presence of geometric dephasing, dissipative decoherence, and technical dilution. The framework derived here separates these contributions for mirror-symmetric Rarity--Tapster matter-wave Bell interferometers.

We show that evaluating the Bell correlation at zero path difference allows us to perform an exact extraction of the momentum-averaged Bell-amplitude without any contribution from geometric dephasing. This framework is then extended to the different mass scenario, and for the  $^{3}$He$^*$/$^{4}$He$^*$ system, the contribution from the known atomic decoherence channels are found to be negligible at current sensitivity. The only non-negligible differential contribution at current sensitivity arises from source-occupancy imbalance, so Eq.\,\eqref{eq:null_benchmark} defines the expected ratio in the absence of additional differential decoherence. A measured deviation beyond this benchmark, once the statistical uncertainty is controlled, would indicate differential Bell-amplitude loss beyond the bounded atomic-physics channels identified in this work.

Although the numerical estimates given here are specialized to metastable helium, the underlying diagnostic logic is more general. The ZPD extraction, SU(2) channel classification, and symmetry-based cancellation of linear gradients apply to mirror-symmetric momentum-entangled Bell interferometers more broadly, with platform dependence entering through the relevant mass, recoil, magnetic, and source-statistics parameters. These results establish the systematic framework required to interpret Bell-amplitude deficits in current matter-wave Bell experiments and a platform-independent tool to identify which systematic constraints remain negligible and which become design-limiting as interferometer size and interrogation time increase.

Ultimately, our results provide the foundation needed to interpret current matter-wave experiments and to design future tests looking for decoherence. As future experiments combine large momentum transfers with microgravity to create macroscopic superpositions, they will become sensitive enough to test decoherence models. By providing the strict diagnostic tools needed to isolate these mass-dependent effects, this framework paves the way for future Bell tests to explore the boundary between quantum mechanics and gravity.
\begin{acknowledgments}
We thank Piotr Deuar for carefully reading the manuscript and fruitful discussions. This work was supported through Australian Research Council Discovery Project Grant Nos. DP240101346 and DP240101441. S.S.H. was supported by Australian Research Council Future Fellowship Grant No. FT220100670. S. K. was supported by an Australian Government Research Training Program scholarship.
\end{acknowledgments}

\appendix

\section{Noise channel classification}\label{app:A}
In this appendix, we derive the Lindblad operators and discuss the physical reasoning behind the noise channels in Sec.\,\ref{sec3e}.

\subsection*{A1. Shot-to-shot phase jitter: $\sigma_z$ dephasing channel}
Fluctuations in the Bragg beam phase or ambient magnetic field can impart a random phase $\delta \phi$ to each atom. We can treat them as Gaussian phase noise with zero mean and variance $\sigma_{\phi}^2$. After taking the ensemble average, we can see the exponential decay ($e^{-\sigma_{\phi}^2/2}$) of the off-diagonal elements of the density matrix. This can be expressed using the local jump operators in a Lindblad master equation:
\begin{equation}
    L_X^{\rm jitter} = \sqrt{\Gamma_\phi}\,\sigma_z^{(X)},\qquad X \in \{A,B\},
    \label{eq:app_jump_jitter}
\end{equation}
where $\Gamma_\phi=\sigma_\phi^2 / (2\tau)$ and $\tau$ is the time between experimental shots. 

Since $\sigma_z$ is diagonal in the $\ket{\uparrow}$, $\ket{\downarrow}$ basis, the longitudinal correlation $T_{zz} = \langle\sigma_z^{(A)}\sigma_z^{(B)}\rangle$ remains $+1$. However, the transverse correlations decay to
\begin{equation}
    T_{xx}(t) = -T_{yy}(t) = e^{-2\Gamma_\phi t}.
    \label{eq:app_transverse_decay}
\end{equation}
It is worth noting that static magnetic gradients produce coherent phases that are identical across shots and should not be modeled as a dephasing channel. They are naturally cancelled by the RT mirror symmetry discussed in Appendix.\,\ref{app:d}.

\subsection*{A2. Background gas scattering}
Under the ultrahigh vacuum conditions of the He$^*$ experiment ($\sim10^{-8}$\,mbar), the random momentum kicks imparted by the background gas collision is larger than the momentum width of the detection window. This process removes the atom out of the modes of interest, destroying all bipartite correlations, including $T_{zz}$.

Since the Bell correlation signal $E(\Phi)$ is constructed from normalized coincidence counts, uniform atom loss where no replacement event is detected does not reduce $\langle A\rangle$. The measured Bell amplitude is only affected if the scattered atom causes an accidental coincidence with an uncorrelated partner:
\begin{equation}
    \langle A\rangle_{\rm meas}=\langle A\rangle\,\frac{N_{\rm true}}{N_{\rm true} + N_{\rm acc}},
    \label{eq:app_accidental}
\end{equation}
where $N_{\rm true}$ and $N_{\rm acc}$ are the true and accidental coincidence rates. In the low-occupancy regime $\bar{n} \ll 1$, accidental coincidences are suppressed, and this contribution is negligible under current operating conditions of He$^*$.

\section{Reduction from source statistics}\label{app:c}
After including the source statistics, the measured Bell amplitude for the single- and dual-species configurations are
\begin{align}
    A_4 &= \langle A_{\rm local}\rangle\,f(\bar{n}_4),\label{eq:app_A4}\\
    A_{3-4} &= \langle A_{\rm local} - \delta A\rangle\,f(\bar{n}_{3-4}),
    \label{eq:app_A34}
\end{align}
where $f(\bar{n})$ is the correction for the mean mode occupancy. Here, $\bar{n}_4$ and $\bar{n}_{3-4}$ are the mean mode occupancies in the single- and dual-species runs, respectively. 

To evaluate the ratio of these amplitudes, we parameterise the occupancy imbalance symmetrically:
\begin{equation}
    \bar{n}_{3-4} = \bar{n}+\frac{\Delta\bar{n}}{2},\qquad
    \bar{n}_4 = \bar{n}-\frac{\Delta\bar{n}}{2},
    \label{eq:app_n_sym}
\end{equation}
where $\bar{n}$ is the average occupancy and $\Delta\bar{n} = \bar{n}_{3-4} - \bar{n}_4$. We can then write 
\begin{equation}
    \frac{A_{3-4}}{A_4}=\frac{\langle A_{\rm local} - \delta A\rangle}{\langle A_{\rm local}\rangle}\frac{f(\bar{n}_{3-4})}{f(\bar{n}_4)}.
    \label{eq:app_ratio_exact}
\end{equation}
Expanding the second factor to first order in $\Delta\bar{n}$ gives
\begin{equation}
    \frac{f(\bar{n}_{3-4})}{f(\bar{n}_4)}\approx 1 + \Delta\bar{n}\,\frac{f'(\bar{n})}{f(\bar{n})}.
\end{equation}
Using $f(\bar{n})=\tfrac{1+\bar{n}}{1+3\bar{n}}$\,\cite{lewisswan2015,wasak2018}, the bell amplitude ratio becomes
\begin{equation}
    \frac{A_{3-4}}{A_4} \approx 1-\frac{\langle\delta A\rangle}{\langle A_{\rm local}\rangle}-\frac{2\,\Delta\bar{n}}{(1+\bar{n})(1+3\bar{n})}.
    \label{eq:ratio_expanded_app}
\end{equation}

\section{Magnetic systematics}\label{app:d}
For a magnetic field with a linear gradient and an atom in a magnetically sensitive state, the interferometric phase acquired by each atom is proportional to the product of the gradient and the arm separation. In the RT geometry, we can see that this linear gradient contribution cancels: 
\begin{equation}
    \delta\varphi_1(B')+\delta\varphi_2(B')
    \propto
    B'(+d)+B'(-d)=0,
    \label{eq:linear_cancel_main}
\end{equation}
where $d$ is the arm separation. This cancellation is a direct consequence of the mirror symmetry of the interferometer at the ZPD point. Thus, constant fields and fields with linear gradient have no effect on the Bell-correlation amplitude, unlike the case for spin\,\cite{shin2019bell}. 

While $^{4}$He$^*$ atoms can be shielded from first-order Zeeman shifts by using the $m_J=0$ sublevel\,\cite{athreya2026bell}, there is no magnetically insensitive sublevel available for $^{3}$He$^*$ ($I=1/2$, $J=1$) and hence every sublevel carries a first-order Zeeman sensitivity. The RT geometry cancels these linear shifts, but this holds only at ZPD. 

Any surviving magnetic dephasing arises from field curvature ($B(z)=\frac{1}{2}B'' z^2$). This effect generates a phase proportional to the atom's center-of-mass position, resulting in a momentum-dependent phase shift:
\begin{equation}
    \delta\varphi_{\mathrm{curv}}(p_A)=\frac{2\,\Delta\mu\,B''\,v_r\,T^3}{3\hbar m}\,p_A,
    \label{eq:phi_curv_main}
\end{equation}
where $\Delta\mu$ is the differential magnetic moment between the two species and $T$ is the mirror-pulse timing. Note that since this phase depends on $p_A$, it is not removed by the ZPD condition. This will reduce the corresponding Bell-amplitude after averaging over the source momentum distribution. 

For the helium parameters of Ref.\,\cite{athreya2026bell,yan2025proposal}, this curvature-induced reduction remains far below current experimental sensitivity for typical magnetic curvatures. Thus, the RT geometry removes the leading magnetic-gradient systematic exactly, while the surviving curvature contribution is small at current operating conditions. Further details are given below.
\subsection*{C1. Linear-gradient cancellation and source center-of-mass drift}
In this section, we show that a residual drift of the center-of-mass (COM) of the source will not break the linear-gradient cancellation of the RT geometry. In the lab frame, the atomic trajectories are uniformly shifted by this drift:
\begin{equation}
    z_1(t)\approx v_{cm} t + v_r t, \qquad z_2(t)\approx v_{cm} t-v_r t.
\end{equation}
However, the interferometric phase depends only on the relative arm separation. Because of this, the single-particle phases remain equal and opposite,
\begin{equation}
    \delta\varphi_1=+\frac{\mu B'}{m}k_{\mathrm{eff}}T^2,\qquad\delta\varphi_2=-\frac{\mu B'}{m}k_{\mathrm{eff}}T^2,
\end{equation}
ensuring that $\delta\varphi_1+\delta\varphi_2=0$ for any $v_{cm}$.

This holds for magnetic field curvature also. If we write the COM trajectories as 
\begin{equation}
    z_{\mathrm{CM},1}(t)=\left(v_{cm}+\frac{p_A}{m}\right)t,
    \qquad
    z_{\mathrm{CM},2}(t)=\left(v_{cm}-\frac{p_A}{m}\right)t,
\end{equation}
and their arm separations as
\begin{equation}
    d_1(t)=+v_r t,\qquad d_2(t)=-v_r t,
\end{equation}
the total Bell phase sum from the curvature cross term becomes
\begin{align}
   & \sum \delta\varphi_{B''}\propto\int B''\Big[d_1(t)z_{\mathrm{CM},1}(t)+d_2(t)z_{\mathrm{CM},2}(t)\Big]dt\notag\\
    &=\int B''\Big[(v_r t)\left(v_{cm}t+\frac{p_A}{m}t\right)-(v_r t)\left(v_{cm}t-\frac{p_A}{m}t\right)\Big]dt\notag\\
    &=\int 2B''v_r\frac{p_A}{m}t^2\,dt.
\end{align}
The $v_{cm}$ term is canceled. As a result, the RT Bell phase is completely immune to any initial source drift through second order field curvature. 
\subsection*{C2. Derivation of the curvature phase}
For the dual-species interferometer, atom $A$ (magnetic moment $\mu_A = \Delta\mu$, momentum $+p_A$ along $\hat{k}_{\rm eff}$) and atom $B$ (magnetic moment $\mu_B = 0$) traverse a field with curvature $B(z)=\frac{1}{2}B''z^2$. Setting the source at origin, the trajectory of atom $A$ along $\hat{k}_{\rm eff}$ is
\begin{equation}
    z_A(t) = \begin{cases}
    \dfrac{p_A}{m}t, & 0 < t < T, \\[6pt]
    \dfrac{p_A}{m}(2T-t), & T < t < 2T,
    \end{cases}
    \label{eq:traj_A}
\end{equation}
where the $\pi$-pulse reverses the momentum at $t=T$. Since $\mu_B=0$, only atom $A$ accumulates a magnetic phase:
\begin{equation}
    \phi_A^{B''} = \frac{\Delta\mu B''}{2\hbar}\int_0^{2T}z_A(t)^2\,dt.
    \label{eq:phi_curv_int}
\end{equation}
Computing the integral over both segments gives us a curvature phase of 
\begin{equation}
    \delta\phi_{\rm curv}(p_A) = \frac{\Delta\mu B'' p_A^2 T^3}{3\hbar m^2}.
    \label{eq:phi_curv_exact}
\end{equation}
This phase is quadratic in $p_A$, and we can linearize it around the mean recoil velocity $v_r = \bar{p}/m$. Writing $p_A = mv_r + \delta p$, where $\delta p \ll mv_r$ is the deviation across the source distribution. Substituting to Eq.\,\eqref{eq:phi_curv_exact} gives
\begin{equation}
    \delta\phi_{\rm curv}=\underbrace{\frac{\Delta\mu B'' v_r^2 T^3}{3\hbar}}_{\text{absorbed into }\Phi_0}+\underbrace{\frac{2\Delta\mu B'' v_r T^3}
    {3\hbar m}\delta p}_{\text{linear, causes dephasing}}+\mathcal{O}(\delta p^2).
    \label{eq:phi_curv_expand}
\end{equation}
We retain only the linear term and writing $\delta p \approx p_A-mv_r$:
\begin{equation}
    \delta\phi_{\rm curv}(p_A)\approx \frac{2\Delta\mu B'' v_r T^3}{3\hbar m}
    \,p_A + \text{const.}
    \label{eq:phi_curv_linear}
\end{equation}
\subsection*{C3. Finite source extent}
In reality, the source is not point like and the atoms are emitted from initial positions distributed across a source cloud. Let us consider a source with a finite Thomas-Fermi radius $R_{TF}$. 

For a linear magnetic gradient, an initial spatial offset does not change the relative arm separation. Consequently, the linear-gradient cancellation remains exact for a finite-sized source. For field curvature, the leading term is the cross term proportional to $B'' z_0 z_{\mathrm{CM}}(t)$. This term is odd in $z_0$ and therefore vanishes after averaging over a symmetric source distribution about the trap center. The leading non-zero contribution comes from the variance of the source size:
\begin{equation}
    B''\langle z_0^2\rangle\sim B''R_{\mathrm{TF}}^2/5
\end{equation}
for a Thomas-Fermi profile. Given $R_{\mathrm{TF}}\sim100~\mu\mathrm{m}$ and typical field curvatures, this contribution remains much smaller than the momentum-spread curvature dephasing retained in the main text.

\section{Generality of the framework}\label{app:f}
The methods developed in the main text (SU(2) classification, ZPD extraction, and the RT symmetry arguments) are not unique to metastable helium. They apply to any momentum-entangled Bell interferometers, that is mirror-symmetric, giving a collective phase $\Phi=\varphi_L+\varphi_R$, and whose higher-order source statistics can be parameterized. Our diagnostic framework can be applied simply by substituting the relevant experimental parameters. While the functional scaling of the systematic bounds remains the same across platforms, the experimental difficulty in satisfying these bound will vary widely. 

In large-momentum-transfer (LMT) implementations, the geometric dephasing envelope narrows as the momentum separation increases. This makes the conventional envelope fitting progressively more fragile, while leaving the ZPD extraction of $\langle A\rangle$ unchanged. The same calibration protocol used in the main text remains applicable for locating $\delta x=0$, but the practical importance of doing so increases with momentum splitting. In microgravity or space-based platforms, the same systematic bounds can become numerically much more stringent because of the longer interrogation time. In particular, the spread of the curvature-induced phase grows rapidly with $T$, and long campaigns also tighten the requirement of stable relative source occupancy. The practical implication is that the same systematics that are negligible in the present helium platform can become design-limiting in next-generation LMT or microgravity Bell experiments. In this sense, the framework developed in the main text becomes more relevant as one moves toward larger interferometers and longer interrogation times.

\bibliography{apssamp}
\end{document}